\documentclass[10pt,journal]{IEEEtran}
\usepackage{amsmath,amsfonts}
\usepackage{algorithmic}
\usepackage{algorithm}
\usepackage{array}
\usepackage[caption=false,font=normalsize,labelfont=sf,textfont=sf]{subfig}
\usepackage{stfloats}
\usepackage{url}
\usepackage{verbatim}
\usepackage{graphicx}
\usepackage{cite}
\usepackage{amssymb}
\usepackage{epsfig}
\usepackage{color}
\usepackage{fancybox}
\usepackage{textcomp}
\usepackage{multirow}
\usepackage{setspace}
\usepackage{psfrag}
\usepackage{mathrsfs}
\usepackage{booktabs}
\usepackage{threeparttable}

\def\BibTeX{{\rm B\kern-.05em{\sc i\kern-.025em b}\kern-.08em
    T\kern-.1667em\lower.7ex\hbox{E}\kern-.125emX}}
\usepackage{balance}

    \def\Complex{{\rm\rule[.23ex]{.03em}{1.1ex}\kern-.3em{C}}}

    \newcommand{\be}{\begin{equation}} \newcommand{\ee}{\end{equation}}
    \newcommand{\bea}{\begin{eqnarray}} \newcommand{\eea}{\end{eqnarray}}
    \newcommand{\benum}{\begin{enumerate}} \newcommand{\eenum}{\end{enumerate}}

    \newcommand{\qn}{{\bf n}}

    \newcommand{\qr}{{\bf r}}

    \newcommand{\qu}{{\bf u}}
    \newcommand{\qv}{{\bf v}}
    
    \newcommand{\qx}{{\bf x}}
    \newcommand{\qy}{{\bf y}}
    \newcommand{\qz}{{\bf z}}

    \newcommand{\qF}{{\bf F}}
    \newcommand{\qG}{{\bf G}}
    \newcommand{\qH}{{\bf H}}
    \newcommand{\qI}{{\bf I}}

    \newcommand{\qQ}{{\bf Q}}
    \newcommand{\qR}{{\bf R}}

    \newcommand{\qPhi}{{\boldsymbol \Phi}}

    \newcommand{\calA}{{\cal A}}
    \newcommand{\calB}{{\cal B}}

    \newcommand{\rl}[1]{\color{red}#1}

\begin{document}

\title{Experimental Evaluation of Multiple Active RISs for 5G MIMO Commercial Networks}

\author{Feng-Ji~Chen, Chao-Kai~Wen, \IEEEmembership{Fellow,~IEEE}, and De-Ming~Chian, \IEEEmembership{Member,~IEEE}
\thanks{F.-J.~Chen, C.-K.~Wen, and D.-M.~Chian are with the Institute of Communications Engineering, National Sun Yat-sen University, Kaohsiung 804, Taiwan (e-mail: {\rm king19635@gmail.com}, {\rm chaokai.wen@mail.nsysu.edu.tw}, {\rm icefreeman123@gmail.com}).}
}

\maketitle

\begin{abstract}
While numerous experimental studies have demonstrated the feasibility of reconfigurable intelligent surface (RIS) technology, most have primarily focused on extending coverage. In contrast, this paper presents an experimental evaluation of multiple active RISs deployed in a 5G multiple-input multiple-output (MIMO) commercial network, emphasizing enhancements in channel rank and throughput. We propose a low-complexity, codebook-based beamforming algorithm specifically tailored for multi-RIS configurations, which diversifies directional channels and reduces reliance on explicit channel state information. Field tests using a commercial base station and user equipment reveal that the multi-RIS system can improve channel rank and throughput by up to 14\% compared to single-RIS deployments, while maintaining low computational complexity. These findings underscore the practical benefits of active multi-RIS systems for next-generation networks.
\end{abstract}


\section{Introduction}
Next-generation cellular networks aim to access broader spectrum resources by operating at higher frequencies. However, high-frequency communications suffer from severe path attenuation, requiring higher transmission power and larger base station (BS) antenna arrays to ensure wide coverage and rich scattering. These requirements increase deployment costs and complicate BS design. Reconfigurable intelligent surfaces (RISs) offer a promising solution by dynamically adjusting reflection angles to enhance multipath propagation \cite{CapRIS-2021} and improve coverage \cite{Tang2021RISpathloss}, thereby reducing the need for dense BS deployments. Although early research focused on passive RISs for their low power consumption, the inherent double-fading in reflected channels limits their performance. Active RISs, which amplify signals to counteract multiplicative fading, have demonstrated significant improvements in energy efficiency (EE) and overall performance \cite{ARISsystem-2021}. 

Despite these advances, practical RIS deployment faces several challenges. A major obstacle is the acquisition of accurate channel state information (CSI), which is essential for RIS beamforming. To address this issue, several blind beamforming algorithms \cite{CSM-2023,Chian-2024} have been proposed that rely on probing measurements to infer optimal beamformers without explicit CSI estimation. However, these algorithms involve a trade-off between performance and the number of measurements or iterations required. Moreover, the optimization range of individual RISs is inherently limited, prompting growing interest in multi-RIS systems. Compared to single-RIS configurations, multi-RIS deployments can more efficiently extend optimization coverage, serve multiple users \cite{Ag2023UEmRIS}, and create a richer scattering environment through diverse reflected channels, enhancing throughput and improving effective channel rank \cite{Do2021MRIS,MInfMRIS-2023}. Furthermore, appropriately deploying multiple RISs based on the environment can maximize throughput \cite{Xu2024RIS}. Nevertheless, controlling multiple RISs poses additional challenges compared to single-RIS setups.

Numerous experimental studies have demonstrated the feasibility of RIS technology and proposed practical methods for real-world implementation \cite{MRISblind-2023,CovEnhanceRIS-2024,FTRIS-2024,RFmagus-2024}. However, most experimental efforts have focused on extending service coverage or enhancing EE, with little attention given to validating RIS capabilities for improving channel rank in practical settings. In fact, multi-rank multiple-input multiple-output (MIMO) communication is essential for achieving extremely high throughput. Although theoretical analyses indicate that multi-RIS deployments are more effective than single-RIS setups in enhancing channel rank, experimental validation of this concept has been lacking due to the complexities involved in controlling multiple RISs.

This paper presents the first experimental evaluation of a multi-RIS system deployed in a commercial 5G MIMO network for rank improvement. A low-complexity, codebook-based beamforming algorithm is introduced, optimized for multi-RIS configurations to enhance directional diversity while minimizing reliance on explicit CSI. The algorithm converges in a limited number of iterations and is fully implementable on commercial user equipments (UEs). Field tests using a commercial BS and UE demonstrate that the multi-RIS system can improve channel rank and throughput by up to 14\% compared to single-RIS deployments, all while maintaining low computational complexity. The results highlight the practical benefits of active multi-RIS systems for enhancing MIMO performance in current 5G and future 6G networks, offering an effective solution to the challenges posed by high-frequency communications.

\section{System Model and Performance Metrics}

\subsection{System Model} 
A 5G NR-compatible OFDM system is considered, where the BS and UE are equipped with $N_{\rm t}$ and $N_{\rm r}$ antennas, respectively. The system deploys $L$ active RIS arrays, with the ${ \ell}$-th RIS comprising $K_{ \ell}$ elements. Since double reflections among multiple RISs are typically more than 10 dB weaker than single reflections in real-world scenarios, channels involving multiple reflections are neglected. The received signal per subcarrier at the MIMO receiver is modeled as
\begin{equation}\label{key}
\qy =
\underbrace{ \left(\qH_{\rm d} +
\sum_{\ell=1}^{L} \qG_{ \ell} \qPhi_{ \ell} \qF_{ \ell} \right) \sqrt{P_{\rm t}} }_{ \triangleq \qH }
\qx +
\underbrace{
\sum_{\ell=1}^{L} \qG_{ \ell} \qPhi_{ \ell} \qz_{ \ell} +\qv }_{ \triangleq \qn},
\end{equation}
where $\qx \in \mathbb{C}^{N_{\rm t}}$ is the transmitted signal with unit norm ${\Vert \qx \Vert}^2 = 1$, and $P_{\rm t}$ is the per-antenna transmit power. Here, 
$\qH_{\rm d} \in \mathbb{C}^{N_{\rm r} \times N_{\rm t}}$ represents the direct BS-UE channel, while 
$\qG_{\ell} \in \mathbb{C}^{N_{\rm r} \times K_{\ell}}$ and $\qF_{\ell} \in \mathbb{C}^{K_{\ell} \times N_{\rm t}}$ denote the RIS-UE and BS-RIS channels, respectively, for the $\ell$-th RIS. The diagonal matrix 
$\qPhi_{ \ell} \in \mathbb{C}^{K_{ \ell} \times K_{ \ell}}$ characterizes the RIS element gains and phase shifts. The noise term $\qn$ includes both RIS noise $\qz_{\rl \ell}$ and thermal noise $\qv$ \cite{Chian-2024}.
The gains and phase shifts in $\qPhi_{ \ell}$ are implemented by low-noise amplifiers (LNAs) and phase shifters (PSs) as
\begin{equation}
\qPhi_{ \ell} = c \cdot \qQ_{ \ell},
\end{equation}
where $c$ is the constant LNA gain and
$\qQ_{ \ell} = {\rm diag} (e^{j\theta_{\ell,1}}, \dots , e^{j\theta_{\ell,K_{ \ell}}}) \in \mathbb{C}^{K_{ \ell} \times K_{ \ell}}$ represents the RIS phase shifts. A $B$-bit PS uniformly quantizes the phase over $[0,2\pi)$ with a quantization step $\Delta\theta = 2\pi / 2^B$.

\subsection{Performance Metrics}

We evaluate the multi-RIS design using the following metrics:

\emph{Channel Capacity}---The theoretical maximum transmission rate is determined by
\begin{equation} \label{eq:Capacity}
C = {\rm log_2}{\rm det}(\qI+\widetilde{\qH}\widetilde{\qH}^H),~~\text{(bits/Hz)}
\end{equation} 
where $\widetilde{\qH} \triangleq \qR_{\rm n}^{-1/2}\qH$ is the whitened channel, and the noise covariance matrix is $\qR_{\rm n} = \mathbb{E}\{\qn\qn^H\}$. Note that $C$ is function of $\{\qQ_{ \ell}\}_{{ \ell}=1}^{L}$; for brevity, the dependence on $\qQ_{ \ell}$ is ommitted.

\emph{Reciprocal Condition Number}---This metric indicates the channel condition, which relates to channel rank and spectral efficiency. It is defined as 
\begin{equation}
\varepsilon = \sigma_{\rm min} / \sigma_{\rm max},
\end{equation}
where $\sigma_{\rm min}$ and $\sigma_{\rm max}$ are the minimum and maximum singular values of $\widetilde{\qH}$, respectively. A higher $\varepsilon$ reflects improved multiplexing capabilities. 

\emph{EE}---Active RISs amplify signals to mitigate path loss at the cost of increased power consumption. Following \cite{Ma-2023}, the total power consumption is given by 
\begin{equation} \label{eq:power consumption}
P_{\rm c} = P_{\rm BS} + \sum_{\ell=1}^{L} P_{{\rm ARIS},\ell},
\end{equation}
where $P_{\rm BS}$ and $P_{{\rm ARIS},\ell}$ denote the power consumptions of the BS and the $\ell$-th active RIS, respectively. The BS power is modeled as 
\begin{equation}
P_{\rm BS} = \upsilon_{\rm BS} N_{\rm t}P_{\rm t} + P_{\rm c,BS}
\end{equation}
where $\upsilon_{\rm BS}$  denotes the reciprocal of the power amplifier efficiency, and $P_{\rm c,BS}$ is the BS hardware consumption. The power consumption of the $\ell$-th active RIS is modeled as
\begin{equation}   
P_{ {\rm ARIS},\ell } = \upsilon_{\rm LNA}
(  P_{\rm t} {\Vert \qPhi_{ \ell} \qF_{ \ell} \qx \Vert}^2 + {\Vert \qPhi_{ \ell} \qz_{ \ell} \Vert}^2  )
+ K_{ \ell}  P_{\rm PS}
+ P_{\rm s,ARIS},
\end{equation}
where $\upsilon_{\rm LNA}$ is the reciprocal of LNA efficiency, $P_{\rm PS}$ denotes the static power of a single PS (dynamic consumption is omitted due to its negligible impact on EE), and $P_{\rm s,ARIS}$ accounts for the control board and other static circuit consumption. The EE is then defined as 
\begin{equation}
{\tt EE} = {{\tt BW} \times C}/{P_{\rm c}}  ~\text{(bits/J)}.
\end{equation}
where ${\tt BW}$ represents the system bandwidth.

\section{Multi-RISs Algorithm}

Optimizing multi-RIS systems is considerably more complex than optimizing single-RIS systems. In a multi-RIS setup, the total number of possible configurations is $2^{BK_1} \times 2^{BK_2} \times \dots \times 2^{BK_L}$, which grows exponentially with the number of RISs. Existing model-free approaches, such as the conditional sample mean (CSM) algorithm \cite{CSM-2023} and the blind greedy (BG) algorithm \cite{Chian-2024}, either require an excessive number of samples or suffer from prohibitive computational complexity. To address these challenges, we propose a codebook-based algorithm.

\emph{Codebook Generation}---The angular codebook is generated offline using steering vectors based on the known antenna locations of each RIS. Specifically, the beamforming vector is defined as 
\begin{equation}
\qv (\alpha,\beta) =   \left[ e^{j \qu^T  \qr_{1}}, e^{j \qu^T  \qr_{2}},...,e^{j \qu^T  \qr_{K}}   \right],
\end{equation}
where $\qu = \frac{2\pi}{\lambda} [\sin\beta \cos\alpha, \, \sin\beta \sin\alpha, \, \cos\beta]^T$ is the wave vector (with wavelength $\lambda$), and $\qr_k = [x_k, y_k, z_k]^T$ denotes the Cartesian coordinates of the $k$-th RIS element relative to its center, which is assumed to lie in the local $x$–$z$ plane (i.e., $y=0$). The azimuth angle $\alpha$ spans $[-180^\circ, 180^\circ]$ (positive from the $x$-axis toward the $y$-axis), and the elevation angle $\beta$ spans $[-90^\circ, 90^\circ]$ (positive upward from the $xy$-plane). 
To implement a $B$-bit digital PS, each entry of $\qv(\alpha,\beta)$ is quantized to the nearest phase level as
\begin{equation}
\tilde{\qv} (\alpha,\beta) =   \left[ e^{j {\tt{Q}}( \qu^T  \qr_1 )} , e^{j {\tt{Q}}( \qu^T  \qr_2)},...,e^{j {\tt{Q}}( \qu^T  \qr_K)}  \right],
\end{equation}
where the quantization function is defined as $
{\tt{Q}}(x) = \Delta\theta \cdot \lfloor \frac{\mod(x, 2\pi)}{\Delta\theta} + 0.5 \rfloor$.

A uniform search over the angular domain is then performed. Let the total number of iterations be $T = M \times N$, where $M$ and $N$ denote the number of discrete steps for the azimuth and elevation angles, respectively. To prevent oversampling, the number of samples is limited by $T \leq 2^{BK}$. The search sets are defined as 
\begin{subequations} \label{eq_codebook}
\begin{align}
\calA &= \left\{ \alpha
\,\big|\,
\alpha = \alpha_0 + \frac{m\pi}{M'},
\, m \ { =} \ 0,1, \dots, M-1 \right\}, \\
\calB &= \left\{ \beta
\,\big|\,
\beta = \beta_0 + \frac{n\pi}{N'},
\, n \ { =} \ 0,1, \dots, N-1  \right\},
\end{align} 
\end{subequations}
where $\alpha_0$ and $\beta_0$ are the starting angles, and $M'$ and $N'$ determine the angular resolution. To ensure the search remains within the RIS field-of-view, assumed to be at most $\pi$ in each dimension, the design satisfies $M \leq M'+1$ and $N \leq N'+1$. Notably, the codebook size depends on the angular resolution rather than the number of RIS elements or the number of quantization levels, unlike conventional model-free methods.

\emph{Sequential Search}---Although the codebook reduces the search space to $T$ candidates, jointly optimizing the phase configurations of all RISs remains computationally intensive. To alleviate this, a sequential optimization strategy is adopted, in which each RIS is optimized individually while keeping the others fixed. Channel capacity is employed as the optimization criterion, as it effectively captures both signal amplification and channel rank enhancement.
For the ${ \ell}$-th RIS, the optimal phase configuration is determined as
\begin{equation} \label{eq:optim}
\qQ_{\ell}^{\star} = \mathop{\arg\max}
_{\alpha \in  \calA, \,
\beta \in  \calB }
\ \
C(\qQ_{\ell} (\alpha, \beta)),
\end{equation}
where $ \qQ_{\rl \ell} (\alpha, \beta) = {\rm diag} (\tilde{\qv}  (\alpha, \beta) ) $ and $C(\cdot)$ denotes the channel capacity defined in \eqref{eq:Capacity}. The sequential search across all $L$ RISs results in an overall computational complexity of $\mathcal{O}(LT)$.

\section{Performance Evaluation}

\subsection{Numerical Simulations}

\begin{figure}
    \centering
    \resizebox{2.5 in}{!}{%
    \includegraphics*{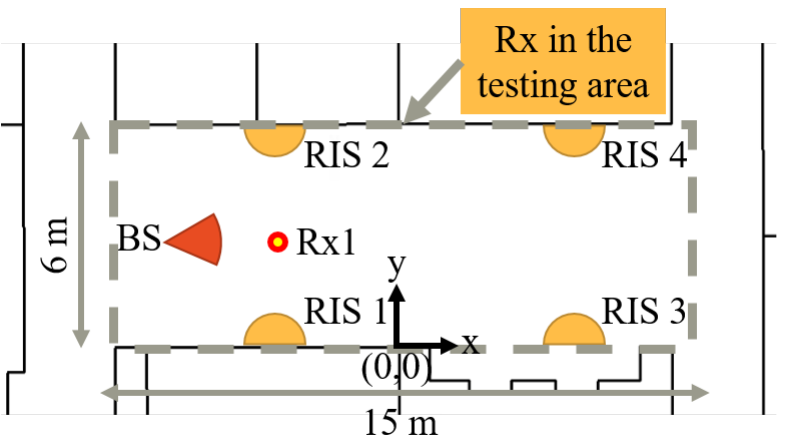} }%
    \vspace{-0.25cm}
    \caption{Simulation environment.}\label{fig:simulation_environment}
    \vspace{-0.25cm}
\end{figure}

Simulations are conducted using Wireless InSite, a ray-tracing software, to extract multipath parameters including path loss, delay, angle of departure, and angle of arrival. These parameters are then used to construct channels based on a 3GPP model replicating the corridor open-space environment of the National Sun Yat-sen University (NSYSU) Electrical and Computer Engineering Building, as illustrated in Fig.~\ref{fig:simulation_environment}. The 8R4T MIMO system settings follow those in \cite{Chian-2024}, except that isotropic antennas are employed at the UE. Four $4\times4$ RIS panels are deployed at the following locations and orientations: RIS1 and RIS3 are positioned at $(-3, 0.3, 2)$ and $(2.3, 0.3, 2)$, respectively, with surface normals pointing along the $+y$-axis; RIS2 and RIS4 are located at $(-3, 5.5, 2)$ and $(2.3, 5.5, 2)$, respectively, with normals pointing along the $-y$-axis.

\begin{figure}
    \centering
    \resizebox{3.25 in}{!}{
    \includegraphics*{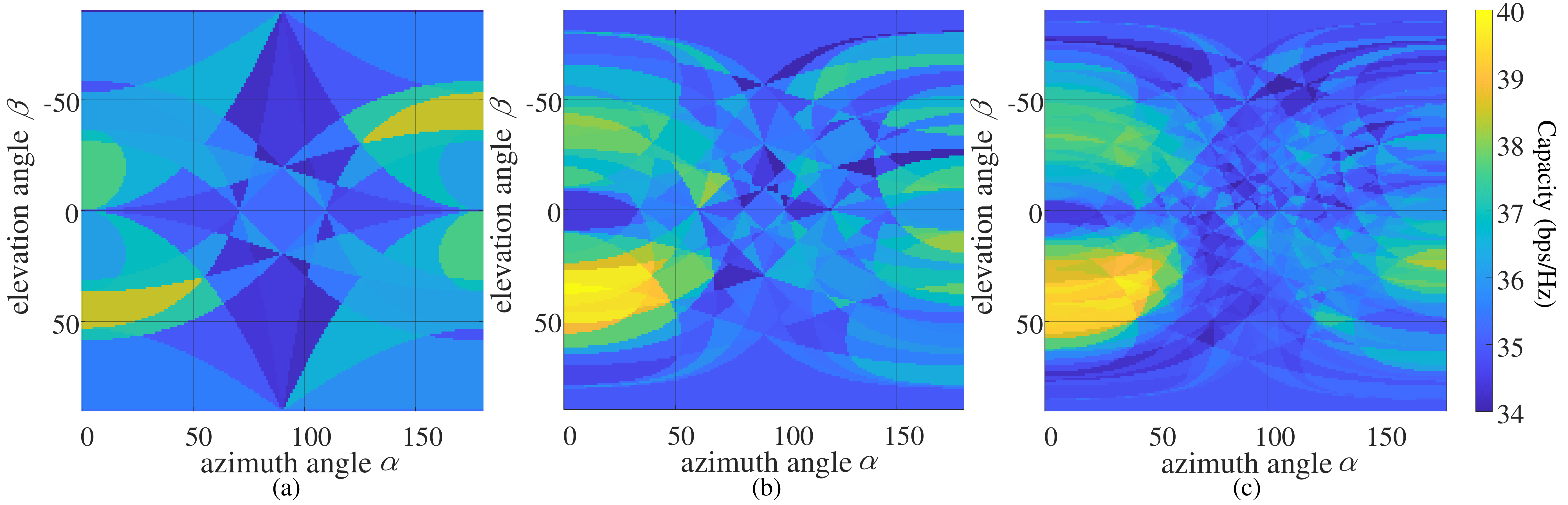} }
    \vspace{-0.25cm}
    \caption{Channel capacity heatmaps for a single RIS at RIS1 with (a) 1-bit, (b) 2-bit, and (c) 3-bit PSs.
    \label{fig:fusion_with_diff_bits}}
    \vspace{-0.25cm}
\end{figure}
 
Fig.~\ref{fig:fusion_with_diff_bits} evaluates the codebook-based algorithm in a single-RIS MIMO system, where the UE is placed at $(-2.7, 3)$ and only RIS1 is activated. An exhaustive search is performed over beamforming points with elevation angles ranging from $-90^\circ$ to $90^\circ$ and azimuth angles from $-180^\circ$ to $180^\circ$. The heatmaps reveal the following: 
 
\emph{1-bit PS}---The maximum channel capacity is approximately 38 bps/Hz. Due to coarse angular resolution, the heatmap exhibits discontinuities and a symmetric pattern centered near $(90^\circ, 0^\circ)$, a result of the limited phase options (i.e., $0^\circ$ and $180^\circ$). This observation implies the need for higher phase resolution to ensure directional precision.

\emph{2-bit and 3-bit PS}---With 2-bit and 3-bit PSs, the maximum channel capacity reaches approximately 39 bps/Hz, approaching the ideal $\infty$-bit performance. The heatmaps reveal distinct hot spots and smoother transitions, indicating that higher phase resolutions improve angular accuracy and mitigate angular ambiguity.

\begin{figure}
    \centering
    \resizebox{2.5 in}{!}{
    \includegraphics*{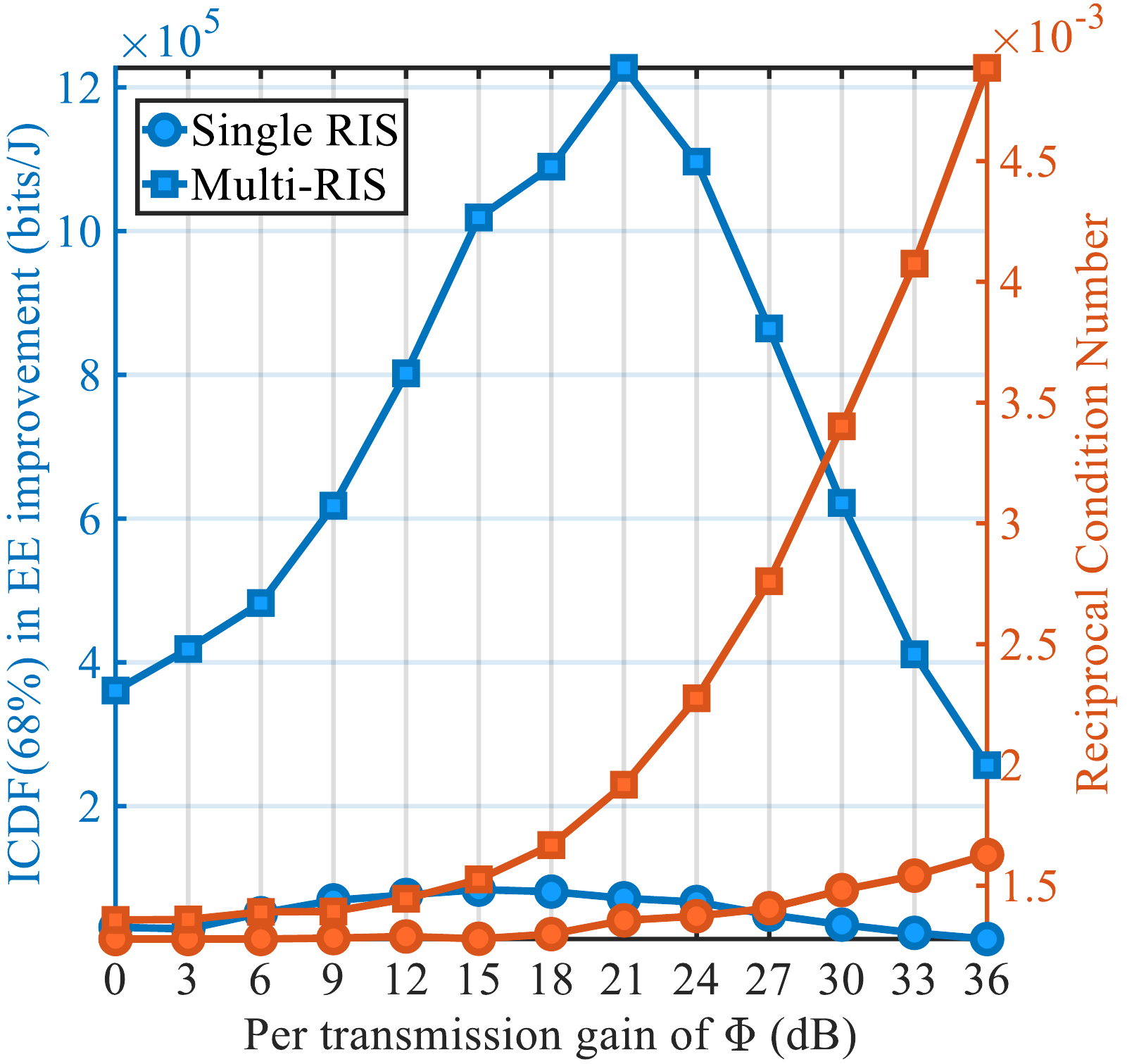} }
    \caption{Comparison of EE and channel rank for single-RIS and multi-RIS systems.} \label{fig:Sigle_versus_Multi}
    \vspace{-0.5cm}
\end{figure}

Next, we compare EE and channel rank between two RIS deployment strategies while maintaining the same total number of RIS elements: (i) a single $8\times8$ RIS located at RIS1, and (ii) four $4\times4$ RIS panels deployed at RIS1, RIS2, RIS3, and RIS4. EE is evaluated using the inverse cumulative distribution function (ICDF at 68\%) as in \cite{Chian-2024}, while the reciprocal condition number serves as a proxy for channel rank improvement. The codebook-based algorithm scans the codebooks defined in (\ref{eq_codebook}) with parameters $(\alpha_0, \beta_0, M', N', M, N) = \left(0, -\frac{\pi}{2}, 3, 3, 4, 4\right)$. Power consumption parameters are set as $P_{\rm t} = 1\,\text{W}$, $P_{\rm PS} = 0.001\,\text{mW}$, $P_{\rm LNA} = 600\,\text{mW}$, $\upsilon_{\rm BS} = \upsilon_{\rm LNA} = 3$, $P_{\rm c,BS} = 6\,\text{W}$, and $P_{\rm s,ARIS} = 0.48\,\text{W}$.

As shown in Fig.~\ref{fig:Sigle_versus_Multi} (dual Y-axis plot), the multi-RIS system achieves superior EE and higher channel rank compared to the single-RIS system. In particular, at an amplification factor of 21 dB, the multi-RIS configuration achieves over a tenfold improvement in EE. Furthermore, the multi-RIS deployment creates richer directional diversity, thereby enhancing effective channel rank and spectral efficiency.

\begin{figure}
    \centering
    \resizebox{3.0 in}{!}{%
    \includegraphics*{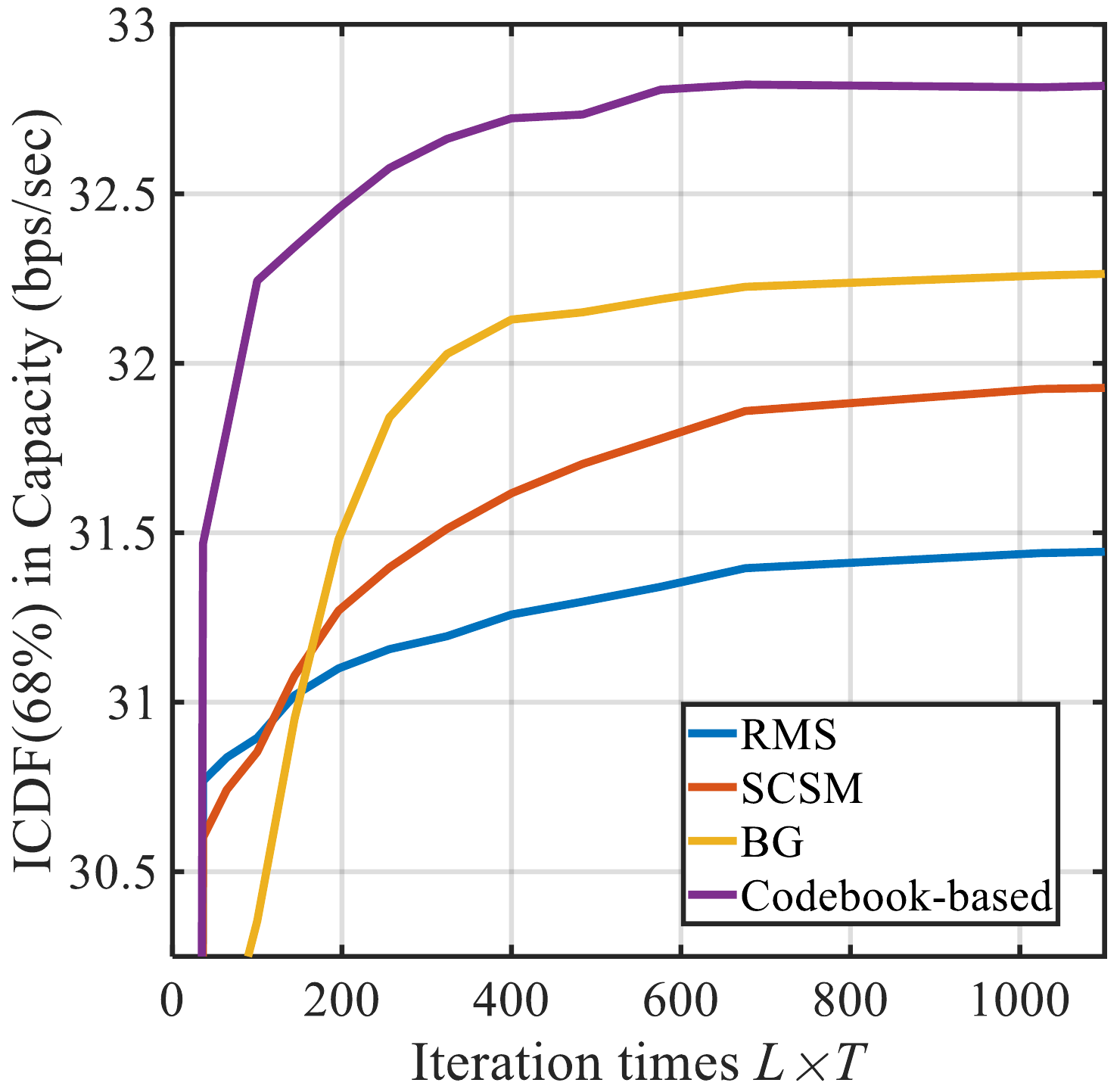} }%
    \vspace{-0.25cm}
    \caption{Iteration count versus capacity: Comparison among RMS, SCSM \cite{MRISblind-2023}, BG \cite{Chian-2024}, and the proposed codebook-based algorithm.} \label{fig:diff_algorithm}
    \vspace{-0.25cm}
\end{figure}

Fig.~\ref{fig:diff_algorithm} compares the computational complexity and capacity performance of four blind beamforming algorithms in a multi-RIS MIMO system with four $4\times4$ RIS panels, optimized sequentially in the order of RIS1, RIS2, RIS3, and RIS4. All algorithms employ a 2-bit PS resolution. The total number of search points is defined as $T = M^2 = N^2$ with $M = N = 2, 3, \ldots, 19$; for the codebook-based method, $M'$ and $N'$ are set to $\sqrt{T} - 1$. The compared algorithms are: 
\begin{itemize} 
\item \textbf{Random Maximum Sampling (RMS):} Selects the best solution from $T$ random samples per RIS panel. 

\item \textbf{Sequential CSM (SCSM) \cite{MRISblind-2023}:} Attempts to find statistically optimal solutions through extensive sampling (same $T$ as RMS).

\item \textbf{BG \cite{Chian-2024}:} A two-stage algorithm that combines random sampling with a 2-bit greedy search. If $T \leq (2^B-1) \times K_{ \ell}$, only random sampling is performed; if $T > (2^B-1) \times K_{ \ell}$, the method allocates $T_{\rm Random} = T-T_{\rm Greedy}$ random samples and  $T_{\rm Greedy} = (2^B-1) \times K_{ \ell}$ samples for the greedy search.

\end{itemize}

All algorithms traverse each RIS only once. The results demonstrate that the proposed codebook-based algorithm achieves the highest capacity with the fewest iterations. When the total iteration count (i.e., $L \times T $) is below 192, RMS performs second-best, whereas BG and SCSM suffer from insufficient sample sizes for effective convergence. For $L \times T  \geq 192$, BG benefits from its second-stage refinement, although its performance remains limited by initial random sampling. SCSM approaches the statistically optimal solution but requires up to $1000 \cdot L$ samples, which exceeds the maximum number of iterations considered in the figure. RMS converges even more slowly than SCSM.

\subsection{Commercial Field Tests}

\begin{figure}[ht]
    \centering
    \resizebox{3 in}{!}{%
    \includegraphics*{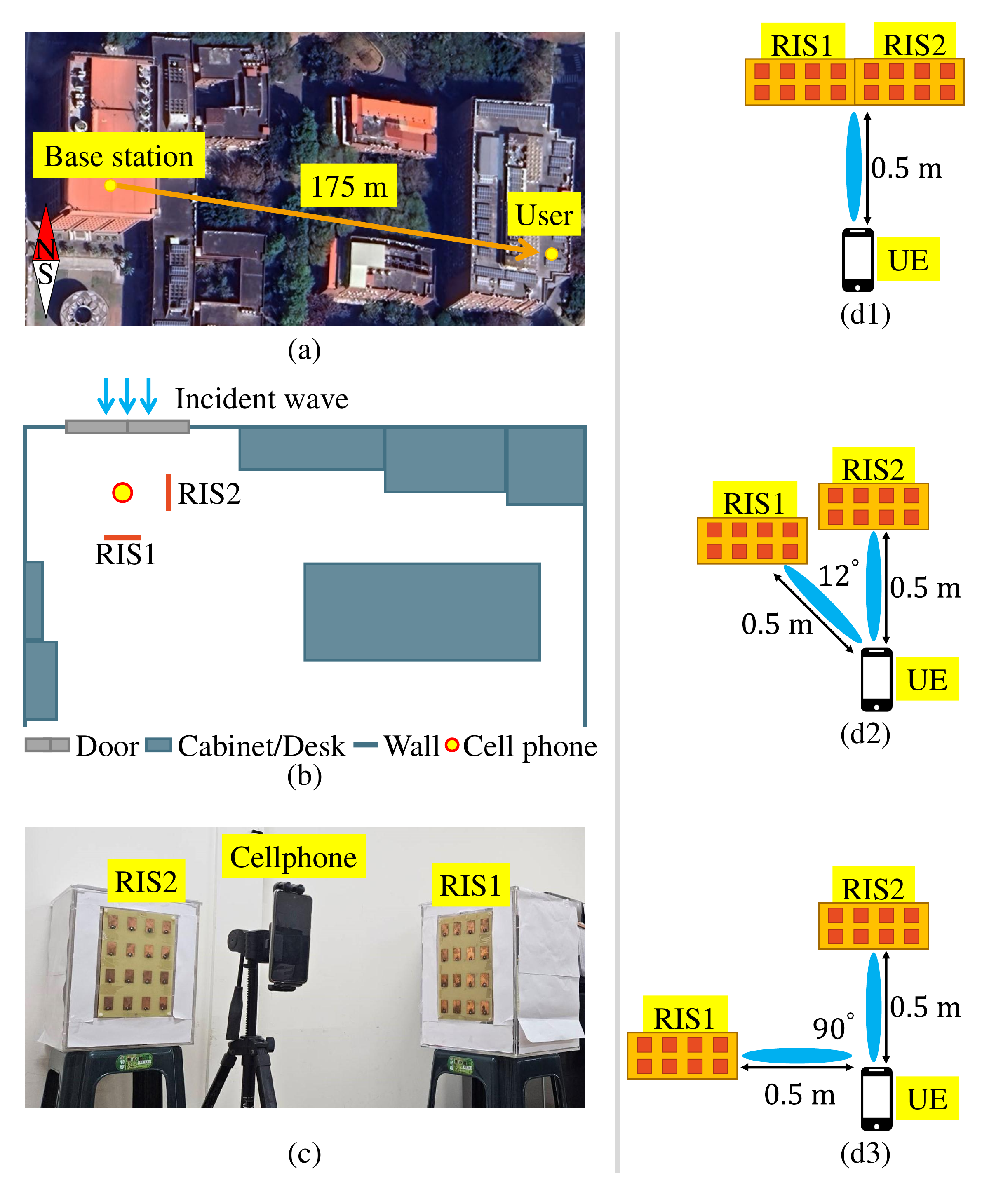} }%
    \vspace{-0.25cm}
    \caption{Commercial testing field: (a) Aerial photograph of an outdoor BS and an indoor office; (b) Indoor testing room plan; (c) Field testing scenario; (d1--d3) Three different measurement scenarios.} \label{fig:Measurement_scanrio}
    \vspace{-0.25cm}
\end{figure}

\begin{table*}
    \begin{center}
    \begin{footnotesize}
    \begin{threeparttable}
        \caption{Performance Comparison of Different Methods in Commercial Field Tests}
        \label{tab:Performance_table}

        \begin{tabular}{|c||c|c|c|c|c|c|c|c|c|c|c|c|c|c|c|}
            \hline
            \multirow{2}{*}{} &
            \multicolumn{3}{c|}{RSRP (dBm)} &
            \multicolumn{3}{c|}{SINR (dB)} &
            \multicolumn{3}{c|}{RI} &
            \multicolumn{3}{c|}{CQI} &
            \multicolumn{3}{c|}{Max. Throughput (bps)} \\
            \cline{2-16}
            {} &
            d1 & d2 & d3 &
            d1 & d2 & d3 &
            d1 & d2 & d3 &
            d1 & d2 & d3 &
            d1 & d2 & d3 \\
            \hline
            \hline
            w/o RIS & 
            \multicolumn{3}{c|}{-94} &
            \multicolumn{3}{c|}{13} &
            \multicolumn{3}{c|}{2} &
            \multicolumn{3}{c|}{13} &
            \multicolumn{3}{c|}{877.8} \\
            \hline
            RMS &
            -91 & -91 & -93 &
            17 & 17 & 14 &
            2 & 2 & 2 &
            13 & 13 & 13 &
            1133 & 1104 & 1104 \\
            \hline
            SCSM &
            -91 & -89 & -93 &
            17 & 17 & 12 &
            2 & 2 & 2 &
            11 & 13 & 13 &
            1104 & 1133 & 1048 \\
            \hline
            BG &
            -91 & -90 & -95 &
            18 & 18 & 12 &
            2 & 2 & 2 &
            13 & 13 & 13 &
            1133 & 1133 & 1133 \\
            \hline
            Codebook-Based &
            -86 & -89 & \textbf{-90} &
            21 & 19 & \textbf{16} &
            2 & 2 & \textbf{3} &
            13 & 12 & \textbf{9} &
            1133 & 1104 & \textbf{1274} \\
            \hline
        \end{tabular}
    \end{threeparttable}
    \end{footnotesize}
    \end{center}
    \vspace{-0.5cm}
\end{table*}

\subsubsection{Test Setup}

As shown in Fig.~\ref{fig:Measurement_scanrio}(a), a commercial BS operated by Chunghwa Telecom Co., Ltd. is located on the rooftop of the NSYSU library. The BS transmits 5G signals with up to rank-4 spatial multiplexing in the 3.420--3.510 GHz band over a 90 MHz bandwidth, with a maximum transmit power of 300 W. A Xiaomi Note 10 5G smartphone is used as the UE, along with the Network Signal Guru application to measure the desired performance metrics. Specifically, based on measurements obtained from the Synchronization Signal Block (SSB), the reference signal received power (RSRP) and signal-to-interference-plus-noise ratio (SINR) are computed, with SINR adopted as the primary metric for RIS optimization. In addition, MIMO indicators, including the rank indicator (RI) and channel quality indicator (CQI) \cite{3GPP-2021}, are collected to estimate the maximum achievable downlink throughput.

The test setup employs two RIS panels with architectures similar to \cite{Chian-2024} but operating in a different frequency band. Each panel, sized at $17\,\text{cm} \times 26\,\text{cm}$ (as shown in Fig.~\ref{fig:Measurement_scanrio}(c)), operates at the same frequency as the commercial BS and features a $2\times4$ active RIS design. Each element provides an amplification gain of 17 dB.

\subsubsection{Measurement Scenarios}
Fig.~\ref{fig:Measurement_scanrio}(b) shows the overall test environment. In this setup, the signal transmitted from the outdoor BS penetrates a large glass door into an indoor laboratory on the 9th floor of the EC Building. Two RIS panels are deployed approximately one meter from the glass door to enhance the received signal, as depicted in Fig.~\ref{fig:Measurement_scanrio}(c). Three measurement scenarios (d1, d2, and d3) were designed to compare multi-RIS configurations against a single-RIS baseline while maintaining a constant total number of RIS elements: 
\begin{itemize}
    \item \textbf{d1}: RIS1 and RIS2 are combined to form a single large $2\times8$ RIS placed 0.5 m from the UE, aligned with the incident wave to deliver a strong focused beam.
    \item \textbf{d2}: RIS1 remains fixed, while RIS2 is shifted along a 0.5~m radius, forming a $12^\circ$ angle between them.
    \item \textbf{d3}: RIS1 remains fixed, while RIS2 is repositioned along the same radius to form a $90^\circ$ angle relative to RIS1.
\end{itemize}

\subsubsection{Performance Comparison}

Table~\ref{tab:Performance_table} summarizes the performance of the different algorithms across the three scenarios. Metrics include RSRP, SINR, RI, CQI, and maximum throughput. For comparison, measurements without RIS deployment are also provided.
Four algorithms were evaluated under a fixed sampling budget of $L \times T  = 2 \times 18$ samples, using identical hardware settings. To maintain a manageable candidate space, different phase resolutions are adopted across the algorithms:
\begin{itemize}
    \item RMS and SCSM: Phase resolution of $B=2$ (as in \cite{MRISblind-2023}), with 18 sampling iterations per RIS panel.
    \item BG: Phase resolution of $B=1$, with 10 random samples and 8 greedy search iterations per RIS panel. 
    \item \textbf{Codebook-based}: Search space limited to $[-30^\circ, 30^\circ]$ in  azimuth, with angular resolutions of $12^\circ$ (azimuth) and $10^\circ$ (elevation), yielding $M=6$ and $N=3$. Phase resolution is set to $B=3$, as the search time is independent of the phase resolution.
\end{itemize}
The field test results reveal the following key observations:

\emph{Signal Enhancement}: Scenario d1 consistently achieves the highest RSRP and SINR across all evaluated methods, delivering up to an 8 dB improvement compared to the no-RIS baseline. In contrast, scenarios d2 and d3 exhibit reduced beamforming efficiency due to the misalignment between RIS panels.

\emph{Algorithm Comparison}: The codebook-based algorithm outperforms RMS, SCSM, and BG in terms of SINR, achieving an additional 4 dB gain in d1. This superior performance is attributed to its more precise angular control and effective mitigation of channel fluctuations, which are particularly critical in dynamic environments.

\emph{Throughput and MIMO Indicators}: Although scenario d1 provides the strongest signal quality, scenario d3 yields the highest throughput, achieving a 14\% improvement over d1. This enhancement is primarily due to an increase in the RI from 2 to 3, enabling an additional spatial stream. Despite a slight CQI degradation resulting from reduced modulation complexity, the overall throughput is significantly improved.

\emph{Multi-RIS Advantage}: The codebook-based method achieves the highest throughput in scenario d3 by leveraging sequential search and precise angular steering, whereas the other algorithms suffer from inter-RIS interference effects. Furthermore, blind beamforming methods are more sensitive to channel variations, which are exacerbated in multi-RIS deployments, as adjusting the configuration of one RIS can affect the optimal settings of others. These results highlight the robustness and practical advantages of the codebook-based approach in dynamic multi-RIS environments.

\section{Conclusions}

This work presented a novel codebook-based algorithm for multi-RIS MIMO systems. Through extensive simulations and commercial field tests, the proposed approach was shown to effectively enhance system performance across several key aspects. First, the multi-RIS deployment significantly improved the channel rank, leading to a 14\% increase in throughput compared to single-RIS configurations, without incurring additional power consumption. Second, the codebook-based design substantially reduced complexity compared to conventional blind methods, enabling faster convergence with fewer iterations and demonstrating strong potential for mobility scenarios. These results, validated through both controlled simulations and practical 5G field tests, confirm the feasibility and advantages of multi-RIS deployments for enhancing current 5G networks. Furthermore, they indicate promising potential for future 6G applications and highlight multi-UE support as an important direction for future research.

\bibliographystyle{IEEEtran}
\bibliography{IEEEabrv, References_Quantized}

\end{document}